\documentclass[runningheads]{CMSIM}

\usepackage{graphicx} % standard LaTeX graphics tool

\usepackage{amsmath,amssymb}

\renewcommand{\thefootnote}

\title*{A Fisher Information Perspective of Relativistic Quantum Mechanics}

\titlerunning{\it Fisher Information Perspective of Relativistic Quantum Mechanics}

\author{
Asher Yahalom
}

\authorrunning{\it Asher Yahalom\inst{1}}

\institute{
Ariel University, Ariel 40700, Israel\\
(E-mail: {\tt asya@ariel.ac.il})
}

\begin{document}
\thispagestyle{empty}
\maketitle
\setlength{\leftskip}{0pt}
\setlength{\headsep}{16pt}
\footnote{
\begin{tabular}{p{11.2cm}r}
\small {\it $16^{th}$CHAOS Conference Proceedings, 13 - 16 June 2023, Heraklion, Crete, Greece} \\
\small C. H. Skiadas (Ed)\\
\small \textcopyright {} 2023 ISAST
%\& \includegraphics[scale=0.35]{CMSIM_Logo}
 \end{tabular}
 }
\begin{abstract}
In previous papers we have shown how Schr\"{o}dinger's equation which includes an electromagnetic field interaction can be deduced from a fluid dynamical Lagrangian of a charged potential flow that interacts with an electromagnetic field. The quantum behaviour was derived from Fisher information terms which were added to the classical Lagrangian. It was thus shown that a quantum mechanical system is drived by information and not only electromagnetic fields.

This program was applied also to Pauli's equations by removing the restriction of potential flow and using the Clebsch formalism. Although the analysis was quite successful there were still terms that did not admit interpretation, some of them can be easily traced to the relativistic Dirac theory. Here we repeat the analysis for a relativistic flow, pointing to a new approach for deriving relativistic quantum mechanics.
\end{abstract}

\newcommand{\beq} {\begin{equation}}
\newcommand{\enq} {\end{equation}}
\newcommand{\ber} {\begin {eqnarray}}
\newcommand{\enr} {\end {eqnarray}}
\newcommand{\eq} {equation}
\newcommand{\eqn} {equation }
\newcommand{\eqs} {equations }
\newcommand{\ens} {equations}
\newcommand {\er}[1] {equation (\ref{#1}) }
\newcommand {\ern}[1] {equation (\ref{#1})}
\newcommand {\ers}[1] {equations (\ref{#1})}
\newcommand {\Er}[1] {Equation (\ref{#1}) }
\newcommand{\br} {\bar{r}}
\newcommand{\tnu} {\tilde{\nu}}
\newcommand{\rhm}  {{\rho \mu}}
\newcommand{\sr}  {{\sigma \rho}}
\newcommand{\bh}  {{\bar h}}
\newcommand {\Sc} {Schr\"{o}dinger}
\newcommand {\SE} {Schr\"{o}dinger equation }
\newcommand {\bR} {{\bf R}}
\newcommand {\bX} {{\bf X}}
\newcommand{\ce}  {continuity equation }
\newcommand{\ces} {continuity equations }
\newcommand{\hje} {Hamilton-Jacobi equation }
\newcommand{\hjes} {Hamilton-Jacobi equations }
\newcommand{\bp}  {\bar{\psi}}
\newcommand{\va}  {{\vec \alpha}}
\newcommand{\mn}  {{\mu \nu}}

\section {Introduction}

Quantum mechanics, is usually interpreted by the Copenhagen school approach. The Copenhagen approach  defies the ontology of the quantum wave function and declares it to be completely epistemological
(a tool for estimating probability of certain measurements) in accordance with
the Kantian \cite{Kant} conception of reality, and its denial of the human ability to grasp any thing "as it is" (ontology). However, historically we also see the development of another school of prominent scholars that interpret quantum mechanics quite differently.  This school believed in the reality of the wave function. In their view the wave function is part of reality much like an electromagnetic field is. This approach that was supported by Einstein and Bohm \cite{Bohm,Holland,DuTe} has resulted in other interpretations of quantum mechanics among them the fluid realization championed by Madelung \cite{Madelung,Complex} which stated that the modulus square  of the wave function is a fluid density and the phase is a potential of the velocity field of the fluid. However, this approach was constrained to wave functions of spin less electrons and could not take into account a complete set of attributes even for slow moving (with respect to the speed of light) electrons.

A non relativistic quantum equation for a spinor was first introduced by Wolfgang Pauli in 1927 \cite{Pauli}. This equation is based on a two dimensional operator matrix Hamiltonian. Two dimensional operator matrix Hamiltonians are currently abundant in the literature (\cite{EYB1} - \cite{EY8}) and describe many types of quantum systems. It is natural to inquire wether such a theory can be given a fluid dynamical interpretation. This question is of great importance as supporters of the non-realistic Copenhagen school of quantum mechanics usually use the spin concept as a proof that nature is inherently quantum and thus have elements without classical analogue or interpretation. A Bohmian analysis of the Pauli equation was given by Holland and others \cite{Holland}, however, the analogy of the Pauli theory to fluid dynamics and the notion of spin vorticity were not considered. This state of affairs was corrected in \cite{Spflu} introducing spin fluid dynamics.

The interpretation of Pauli's spinor in terms of fluid density and velocity variables leads us directly to the nineteenth century seminal work of Clebsch \cite{Clebsch1,Clebsch2} which is strongly related to the variational analysis of fluids. Variational principles for barotropic fluid dynamics are described in the literature. A four function variational principle for an Eulerian barotropic fluid was depicted by Clebsch \cite{Clebsch1,Clebsch2} and much later by Davidov \cite{Davidov} who's main purpose was to quantize fluid dynamics. The work was written in Russian, and was largely unknown in the west. Lagrangian fluid  dynamics (which takes a different approach than Eulerian fluid dynamics) was given a variational description by Eckart \cite{Eckart}.
Ignoring both the work of Clebsch (written in German) and the work of Davidov (written in Russian)  initial attempts in the English written literature to formulate Eulerian fluid dynamics using a variational principle, were given by Herivel \cite{Herivel}, Serrin
\cite{Serrin} and Lin \cite{Lin}. However, the variational principles
developed by the above authors were cumbersome relying on
quite a few "Lagrange multipliers" and auxiliary "potentials". The total number of independent functions in the above
formulations are from eleven to seven, which are much more than the
four functions required for the Eulerian and continuity equations
of a barotropic flow. Thus those methods did not have practical use. Seliger \& Whitham \cite{Seliger} have reintroduced the variational formalism of Clebsch depending on only four
variables for barotropic flow. Lynden-Bell \& Katz \cite{LynanKatz} have described a variational principle in terms of two functions the load $\lambda$ and density $\rho$.
However, their formalism contains an implicit definition for the velocity $\vec v$
such that one is required to solve a partial differential equation in order
to obtain both $\vec v$ in terms of $\rho$ and $\lambda$ as well as its variations.
Much the same criticism holds for their general variational for non-barotropic flows \cite{KatzLyndeb}. Yahalom \& Lynden-Bell \cite{YahLyndeb} overcame the implicity definition limitation by paying the price of adding an additional single variational variable. This formalism allows arbitrary variations (not constrained) and the definition of $\vec v$ is explicit.
The original work of Clebsch and all the following publications assume a non-relativistic fluids in
which the velocity of the flow is much slower than the speed of light in vacuum $c$. This is of course to be expected as the work of Clebsch preceded Einstein's work on special relativity by forty eight years. This can also be based on practical basis as relativistic flows are hardly encountered on earth.

The standard approach to relativistic flows is based on the energy-moment\-um tensor \cite{Padma,Weinberg,MTW}, however, this approach is not rigorous because the definition of an energy-momentum tensor can only be done if a Lagrangian density is provided \cite{Goldstein}. However, no Lagrangian density was known for relativistic flows. In this work we intend to expand Clebsch work to relativistic flow and thus amend this lacuna with a derived Lagrangian density for a relativistic flow from which one can obtain rigorously the energy-momentum tensor of high velocity flows.

A fundamental issue in the fluid interpretation of quantum mechanics still remains. This refers to the meaning of thermodynamic quantities. Thermodynamics concepts like specific enthalpy,
 pressure and temperature are related to the specific internal energy defined by the equation of state as a unique function of entropy and density.
 The internal energy is a part of any Lagrangian density related to fluid dynamics.
 The internal energy  functional can in principle be explained on the basis of the microscopic composition of the fluid using statistical physics. That is the atoms and
 molecules from which the fluid is composed and their interactions impose an equation of state. However, a quantum fluid has no structure
 and yet the equations of both the spin less \cite{Madelung,Complex} and spin \cite{Spflu} quantum fluid dynamics shows that terms analogue
 to internal energies appear. One thus is forces to inquire where do those internal energies originate? Of course one cannot suggest that the
 quantum fluid has a microscopic sub structure as this will defy current empirical evidence suggesting that the electron is a point particle. The answer to this question
 comes from an entirely different scientific discipline known as measurement theory \cite{Fisher,YaFisher,Fisherspin}. Fisher information is a basic notion of measurement theory,
 and is a measure of measurement quality of any quantity. It was demonstrated \cite{Fisherspin} that this notion is  the internal energy of a spin less electron (up to a proportionality constant) and can interpret sum terms of the internal energy of an electron with spin. Here we should mention an attempt to derive most physical theories from Fisher information as described by Frieden \cite{Frieden}. It was suggested  \cite{Fisherspin2} that there exist a velocity field such that the Fisher information will given a complete explanation for the spin fluid internal energy. It was also suggested that one may define comoving scalar fields as in ideal fluid mechanics, however, this was only demonstrated implicitly but not explicitly. A common feature of previous work on the fluid \& Fisher information interpretation of quantum mechanics, is the negligence of electromagnetic interaction thus setting the vector potential to zero. This makes sense as the classical ideal fluids discussed in the literature are not charged. Hence, in order to make the comparison easier to comprehend the vector potential should be neglected. However, one cannot claim a complete description of quantum mechanics lacking a vector potential thus ignoring important quantum phenomena such as the Zeeman effect which depends on a vector potential through the magnetic field, this was taken care of in \cite{Fisherspin3,Fisherspin4}. However, this previous work assumed a non-relativistic flow. In the current paper we  study a relativistic flow and thus suggest a new route leading to relativistic quantum mechanics which is based on a relativistic fluid dynamics with a Lorentz invariant Fisher information term.

 We will begin this paper by introducing a variational principle for a relativistic charged classical particle with a vector potential interaction and a system of the same. This will be followed by the Eckart \cite{Eckart} Lagrangian variational principles generalized for a relativistic charged fluid. We then introduce an Eulerian-Clebsch variational principle for a relativistic charged fluid. Finally the concept of Fisher information will allow us to suggest a new approach to relativistic quantum fluids.

 \section{Trajectories Through Variational Analysis}

We consider a particle travelling in spacetime of a constant metric.
The action ${\cal A}$ of such a particle is:
\beq
{\cal A} = - m c \int  d \tau - e  \int A^\alpha d x_\alpha
\label{Action}
\enq
In the above $\tau$ is the trajectory interval:
\beq
d \tau^2 = \left|\eta^{\alpha \beta} d x_\alpha d x_\beta \right|
= \left| d x_\alpha d x^\alpha \right|
\label{length}
\enq
 $x_\alpha$  are the particle coordinates (the metric raises and lowers indices according to the prevailing custom), $m$ is the particle mass, $e$ is the charge and $A^\alpha$  is the four vector potential  which depend on the particle coordinates.  $A^\alpha$ transforms as a four dimensional vector. Variational analysis results in the following equations of motion:
\beq
m \frac{d u^\alpha}{d \tau}= -\frac{e}{c} u^\beta (\partial_\beta A^\alpha - \partial^\alpha A_\beta), \qquad u^\alpha \equiv \frac{d x^\alpha}{d \tau}, \quad
\partial^\alpha \equiv \frac{\partial}{\partial x_\alpha}, \quad
\partial_\beta \equiv \eta_{\beta \alpha} \partial^\alpha
\label{Eqmot}
\enq
in which the metric $\eta_{\alpha \beta}$ is the Lorentz metric:
\beq
\eta_{\alpha \beta} = \ {\rm diag } \ (1,-1,-1,\--1).
\label{Lorentz}
\enq

\subsection{Partition to Space \& Time}

Given a  space-time with a Lorentz metric the partition into spatial and temporal coordinates is trivial. The spatial coordinates are $\vec x = (x_1,x_2,x_3)$ and the temporal coordinate is $x_0$. As we measure time in the units of seconds which differ from the space units of meters, we introduce $x_0 = c t$, in which $c$ connects the different units. The velocity is defined as:
\beq
\vec v \equiv \frac{d \vec x}{d t}, \qquad v = |\vec v|, \qquad v_\alpha \equiv \frac{d x_\alpha}{d t}= (\vec v,c).
\label{vel}
\enq
In a similar way we dissect $A_\alpha$ into temporal and spatial pieces:
\beq
A_\alpha=(A_0,A_1,A_2,A_3) \equiv  (A_0, \vec A) \equiv (\frac{\phi}{c}, \vec A)
\label{Av}
\enq
the factor $\frac{1}{c}$ in the last term allows us to obtain the equations in MKS units, it
is not needed in other types of unit systems.
Through \ern{Av}, we can define a magnetic field:
\beq
\vec B = \vec \nabla \times \vec A
\label{magnetic}
\enq
($\vec \nabla$ has the standard meaning) and the electric field:
\beq
\vec E =-\frac{\partial \vec A}{\partial t} -\vec \nabla \phi
\label{electric}
\enq
For the subluminal case $v<c$ we may write $d \tau^2$ as:
\beq
d \tau^2 = c^2 dt^2 (1-\frac{v^2}{c^2}), \qquad d \tau = c dt \sqrt{1-\frac{v^2}{c^2}}
= \frac{c dt}{\gamma}, \qquad \gamma \equiv \frac{1}{\sqrt{1-\frac{v^2}{c^2}}}
\label{lengthlsl}
\enq
And using the above equations the spatial piece of \ern{Eqmot} is deduced:
\beq
 \frac{d }{d t} (m \gamma \vec v) = \frac{d }{d t}\left(m \frac{\vec v}{\sqrt{1-\frac{v^2}{c^2}}}\right)= e \left(\vec E + \vec v \times \vec B\right)
\label{Eqmotsl}
\enq

 \subsection{The Lagrangian}

 We may write the action (\ref{Action}) as a temporal integral and thus define a Lagrangian:
 \ber
 {\cal A} &=& \int_{t1}^{t2} L dt, \qquad L = L_0 + L_i
 \nonumber \\
 L_0 &\equiv& -m c \frac{d \tau}{dt}= -\frac{m c^2}{\gamma} = -m c^2 \sqrt{1-\frac{v^2}{c^2}} \simeq  \frac{1}{2} m v^2  - m c^2,
 \nonumber \\
  L_i &\equiv&   - e A^\alpha \frac{d x_\alpha}{d t}=  e(\vec A \cdot \vec v - \phi).
 \label{classparticleact}
 \enr
in the above the $\simeq$ symbol signifies a classical (low speed) approximation. We notice
that the interaction part of the Lagrangian is the same for high and low speeds while the kinetic part takes a different and simpler form for the low speed cases.

\subsection{The Action \& Lagrangian for a System of Particles}

\begin{figure}
\centering
\includegraphics[width= 0.5\columnwidth]{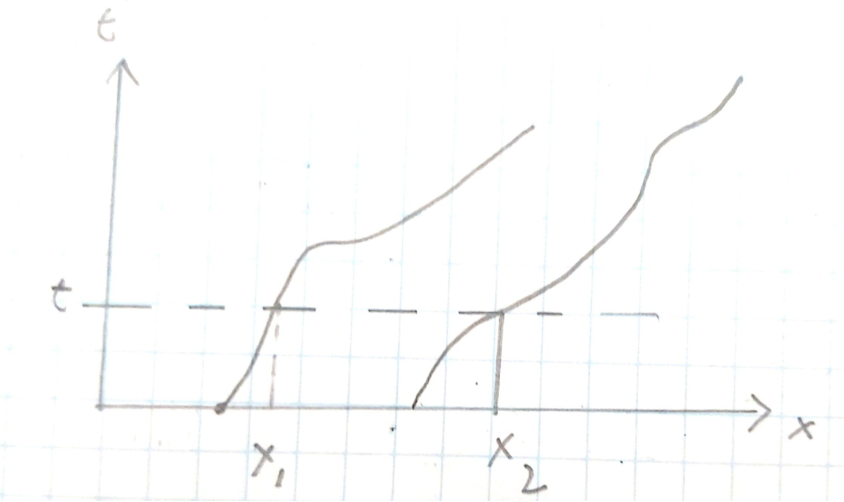}
 \caption{Schematic drawing  of two trajectories piercing a time "plane" which is illustrated as a straight line.}
 \label{timeplane}
\end{figure}

 Consider a system of $N$ particles each with an index $n \in [1-N]$, a corresponding mass $m_n$, charge  $e_n$. Each particle will have a trajectory $x_n^\alpha (\tau_n)$ in which $\tau_n$ measures the interval already propagated along the trajectory. Thus:
 \beq
u_n^\alpha \equiv \frac{d x_n^\alpha}{d \tau_n}.
\label{unal}
\enq
We will assume as usual that the particle trajectories pierce through time "planes", and the "plane" $t$ is pierced at position vector $\vec x_n (t)$, see figure \ref{timeplane} (actually each "plane" is three dimensional).
Thus one can define a velocity $\vec v_n \equiv \frac{d \vec x_n}{dt}$. The action and Lagrangian for each point particle are as before:
\ber
{\cal A}_n &=& - m_n c \int  d \tau_n - e_n  \int A^\alpha (x^\nu_n) d x_{\alpha n}
 = \int_{t1}^{t2} L_n dt, \qquad L_n \equiv L_{0n} + L_{in}
 \nonumber \\
 L_{0n} &\equiv& -\frac{m_n c^2}{\gamma_n}  \simeq \frac{1}{2} m_n v_n^2 -m_n c^2, \quad L_{in} \equiv e_n \left(\vec A (\vec x_n,t) \cdot \vec v_n - \phi (\vec x_n,t)\right).
 \label{classparticlej}
 \enr
 The action and Lagrangian of the system of particles is:
\beq
 {\cal A}_s = \int_{t1}^{t2} L_s dt, \qquad L_s = \sum_{n=1}^{N} L_n.
  \label{classparticlesys}
 \enq
 The variational analysis follows the same lines as for a single particle and we obtain a set of equations of the four dimensional form:
 \beq
m_n \frac{d u_n^\alpha}{d \tau_n}= -\frac{e_n}{c} u_n^\beta (\partial_\beta A_n^\alpha - \partial^\alpha A_{\beta n}), \qquad n \in [1-N].
\label{Eqmotn}
\enq
Or the three dimensional form:
 \beq
 \frac{d}{dt}(\gamma_n \vec v_n) = \frac{e_n}{m_n} \left[\vec v_n \times \vec B (\vec x_n,t) + \vec E(\vec x_n,t) \right], \qquad n \in [1-N].
 \label{equamotionj}
 \enq
in which we do not sum over repeated Latin indices

\section{A Relativistic Charged Fluid - the Lagrangian Approach}

\subsection{The Action and Lagrangian}

The dynamics of the fluid is determined by its composition and the forces acting on it. The fluid is made of "fluid elements" \cite{Eckart,Bertherton}, practically a "fluid element" is a point particle which has an infinitesimal mass $d M_{\vec \alpha}$, infinitesimal charge $d Q_{\vec \alpha}$,  position four vector $x_{\vec \alpha \nu} (\tau_{\vec \alpha})$ and $u_{\vec \alpha \nu} (\tau_{\vec \alpha}) \equiv \frac{d x_{\vec \alpha \nu} (\tau_{\vec \alpha})}{d \tau_{\vec \alpha}}$. Here the continuous vector label $\va$ replaces the discrete index $n$ of the previous section. As the "fluid element" is not truly a point particle it has also an infinitesimal volume $d V_{\vec \alpha}$, infinitesimal entropy $d S_{\vec \alpha}$, and an infinitesimal internal energy $d E_{in~\vec \alpha}$. The action  for each "fluid element" are according to \ern{classparticleact} as follows:
\ber
& & d {\cal A}_{\va} = - dM_{\va} c \int  d \tau_{\va} - d Q_{\va}  \int A^\mu (x^\nu_{\va}) d x_{\mu \va}  +  d {\cal A}_{in~\va},
\nonumber \\
& & d {\cal A}_{in~\va} \equiv -  \int d E_{in~\va} dt.
\label{relaction}
 \enr
The  Lagrangian for each "fluid element" can be derived from the above expression as follows:
\ber
d {\cal A}_{\va} &=& \int_{t1}^{t2} dL_{\va} dt, \qquad
 dL_{\va} \equiv dL_{k \va} + dL_{i \va} - d E_{in~\va}
 \nonumber \\
  d L_{k \va} &\equiv&   -\frac{d M_{\va} c^2}{\gamma_\va}  \simeq \frac{1}{2} d M_{\va}~ v_\va(t)^2  - d M_{\va} c^2
 \nonumber \\
  d L_{i\va} &\equiv& d Q_{\va} \left(\vec A (\vec x_\va (t),t) \cdot \vec v_\va (t) - \phi(\vec x_\va (t),t)\right).
   \label{relLf}
 \enr
all the above quantities are calculated for a specific value of the label $\va$, while the action and Lagrangian of the entire fluid, should be summed (or integrated) over all possible
$\vec \alpha$'s. That is:
\ber
L &=& \int_{\va}  d L_{\vec \alpha}
\nonumber \\
{\cal A} &=& \int_{\va} d {\cal A}_{\vec \alpha} = \int_{\va}  \int_{t1}^{t2}  d L_{\vec \alpha} dt
=   \int_{t1}^{t2}  \int_{\va} d L_{\vec \alpha} dt
 =  \int_{t1}^{t2} L dt.
 \label{fluac1}
 \enr
 It is customary to define densities for the Lagrangian, mass and charge of every fluid element as
 follows:
 \beq
 {\cal L}_{\va} \equiv \frac{d L_{\va}}{d V_{\va}}, \quad
  \rho_{\va} \equiv \frac{d M_{\va}}{d V_{\va}} , \quad
\rho_{c\va} \equiv \frac{d Q_{\va}}{d V_{\va}} , \quad
e_{in~\va} \equiv \frac{d E_{in~\va}}{d V_{\va}}
\label{dens}
\enq
Each of the above quantities may be thought of as a function of the location $\vec x$, where the "fluid element" labelled $\va$ happens to be in time $t$, for example:
\beq
 \rho (\vec x, t) \equiv \rho (\vec x_\va (t), t) \equiv \rho_{\va} (t)
\label{dens2}
\enq
It is also customary to define the specific internal energy $\varepsilon_{\va}$ as follows:
\beq
 \varepsilon_{\va} \equiv \frac{d E_{in~\va}}{d M_{\va}} \quad \Rightarrow  \quad
 \rho_{\va} \varepsilon_{\va} = \frac{d M_{\va}}{d V_{\va}} \frac{d E_{in~\va}}{d M_{\va}} =
 \frac{d E_{in~\va}}{d V_{\va}} = e_{in~\va}
\label{specifint}
\enq
Thus we can write the following equations for the Lagrangian density:
\ber
 {\cal L}_{\va} &=& \frac{d L_{\va}}{d V_{\va}}=
   \frac{{d L}_{k\va}} {d V_{\va}} + \frac{{d L}_{i\va}} {d V_{\va}} -
    \frac{d E_{in~\va}}{d V_{\va}}
 = {\cal L}_{k\va} +  {\cal L}_{i\va} - e_{in~\va}
 \nonumber \\
 {\cal L}_{k\va} &\equiv& -\frac{\rho_{\va} c^2}{\gamma_\va}
 \simeq \frac{1}{2} \rho_{\va} v_\va (t)^2 - \rho_{\va} c^2 ,
  \nonumber \\
  {\cal L}_{i\va} &\equiv& \rho_{c\va} \left(\vec A (\vec x_\va (t),t) \cdot \vec v_\va (t) - \varphi (\vec x_\va (t),t)\right).
 \label{lagdensity}
 \enr
 The above expression allows us to write the Lagrangian as a spatial integral:
 \beq
L = \int_{\va}  d L_{\vec \alpha} = \int_{\va} {\cal L}_{\va} d V_{\va}
= \int {\cal L} (\vec x,t) d^3 x
\label{Lspatint}
\enq
which will be important for later sections of the current paper.

\subsection{Variational Analysis}

Returning now to the variational analysis we introduce the symbols $\Delta \vec x_{\va} \equiv \vec \xi_{\va}$ to indicate a variation of the trajectory $ \vec x_{\va} (t)$ (we reserve the symbol $\delta$  in the fluid context, to a different kind of variation, the Eulerian variation to be described in the next section). Notice that:
\beq
\Delta \vec v_\va (t) =  \Delta \frac{d \vec x_\va (t)}{dt} = \frac{d \Delta \vec x_\va (t)}{dt} = \frac{d \vec \xi_\va (t)}{dt}.
\label{vvar}
\enq
And thus according to \ern{lengthlsl}:
\beq
\Delta \left( \frac{1}{\gamma_\va} \right) = - \frac{\gamma_\va  \vec v_\va(t)}{c^2}
\frac{d \vec \xi_\va (t)}{dt}, \qquad
\Delta \gamma_\va =  \frac{\gamma_\va^3  \vec v_\va(t)}{c^2}
\frac{d \vec \xi_\va (t)}{dt}.
\label{vgam}
\enq
In an ideal fluid the "fluid element" does exchange mass, nor electric charge, nor heat with other fluid elements, so it follows that:
 \beq
\Delta d M_{\vec \alpha} = \Delta d Q_{\vec \alpha} = \Delta d S_{\vec \alpha}  =0.
\label{consvl}
\enq
Moreover, according to thermodynamics a change in the internal energy of a "fluid element" satisfies
the equation in the particle's rest frame:
\beq
\Delta d E_{in~\vec \alpha 0} = T_{\va 0} \Delta dS_{\vec \alpha 0} - P_{\va 0} \Delta dV_{\vec \alpha 0},
\label{thermo}
\enq
the first term describes the heating energy gained by the "fluid element" while the second terms describes the work done by the "fluid element" on neighbouring elements. $T_{\va 0}$ is the temperature of the "fluid element" and $P_{\va 0}$ is the pressure of the same. As the rest mass of the fluid element does not change and does not depend on any specific frame we may divide the above expression by $d M_{\vec \alpha}$ to obtain the variation of the specific energy as follows:
\ber
\Delta \varepsilon_{\va 0}  &=& \Delta\frac{d E_{in~\vec \alpha 0}}{d M_{\va}}
 =  T_{\va 0} \Delta\frac{ dS_{\vec \alpha 0}}{d M_{\va}} - P_{\va 0} \Delta \frac{dV_{\vec \alpha 0}}{d M_{\va}}
 \nonumber \\
 &=& T_{\va 0} \Delta s_{\va 0} - P_{\va 0} \Delta \frac{1}{\rho_{\va 0}}
 = T_{\va 0} \Delta s_{\va 0} + \frac{P_{\va 0}}{\rho_{\va 0}^2} \Delta \rho_{\va 0}. \quad
 s_{\va 0} \equiv \frac{ dS_{\vec \alpha 0}}{d M_{\va}}
\label{thermo2b}
\enr
in which $s_{\va 0}$ is the specific entropy of the fluid element in its rest frame. It follows that:
\beq
\frac{\partial \varepsilon_0 }{\partial s_0} = T_0, \qquad
\frac{\partial \varepsilon_0 }{\partial \rho_0} = \frac{P_0}{\rho_0^2}.
\label{thermo2c}
\enq
Another important thermodynamic quantity that we will use later is the Enthalpy defined for
a fluid element in its rest frame as:
\beq
 dW_{\va 0} = d E_{in~\vec \alpha 0} + P_{\va 0} d V_{\va 0}.
\label{thermo2d}
\enq
and the specific enthalpy:
\beq
 w_{\va 0} =\frac{dW_{\va 0}}{d M_{\va}}= \frac{dE_{in~\vec \alpha 0}}{d M_{\va}}
 +P_{\va 0} \frac{dV_{\va 0}}{d M_{\va}} = \varepsilon_{\va 0} + \frac{P_{\va 0}}{\rho_{\va 0}}.
\label{thermo2e}
\enq
Combining the above result with \ern{thermo2c} it follows that:
\beq
 w_0 = \varepsilon_0 + \frac{P_0}{\rho_0} = \varepsilon_0 + \rho_0 \frac{\partial \varepsilon_0 }{\partial \rho_0}
 = \frac{\partial (\rho_0 \varepsilon_0) }{\partial \rho_0}.
\label{thermo2f}
\enq
Moreover:
\beq
 \frac{\partial w_0 }{\partial \rho_0} = \frac{\partial (\varepsilon_0 + \frac{P_0}{\rho_0})}{\partial \rho_0}
  = - \frac{P_0}{\rho_0^2} + \frac{1}{\rho_0} \frac{\partial P_0}{\partial \rho_0}+ \frac{\partial \varepsilon_0}{\partial \rho_0} =
  - \frac{P_0}{\rho_0^2} + \frac{1}{\rho_0} \frac{\partial P_0}{\partial \rho_0}+ \frac{P_0}{\rho_0^2} = \frac{1}{\rho_0} \frac{\partial P_0 }{\partial \rho_0}.
\label{thermo2g}
\enq
As we assume an ideal fluid, there is no heat conduction or heat radiation, and thus heat can only be moved around along the trajectory of the "fluid elements", that is only convection is taken into account. Thus $\Delta d S_{\vec \alpha 0}  =0$
and we have:
\beq
\Delta d E_{in~\vec \alpha 0} = - P_0 \Delta dV_{\vec \alpha 0}.
\label{thermo2}
\enq
Our next step would to be to evaluate the variation of the volume element. However, before
we do this we establish some relations between the rest frame and any other frame
in which the fluid element is in motion (this frame is sometimes denoted the "laboratory" frame).
First we notice that at the rest frame there is no velocity (by definition), hence according
to \ern{lengthlsl}:
\beq
 d \tau = c dt_0 = c dt \sqrt{1-\frac{v^2}{c^2}} = \frac{c dt}{\gamma}
 \quad \Rightarrow  \quad dt_0 = \frac{dt}{\gamma}.
\label{dt0}
\enq
It is well known that the four volume is Lorentz invariant, hence:
\beq
 dV_0 dt_0 = dV dt = dV dt_0 \gamma, \qquad \Rightarrow dV_0 = \gamma dV .
\label{dV0}
\enq
Thus:
\beq
 \rho_0 = \frac{dM}{dV_0} =   \frac{1}{\gamma}\frac{dM}{dV} = \frac{\rho}{\gamma},
 \qquad \Rightarrow  \quad \rho = \gamma \rho_0.
\label{drho0}
\enq
Moreover, the action given in \ern{relaction} is Lorentz invariant, thus:
\beq
 d E_{in~\va 0} dt_0 = d E_{in~\va} dt = d E_{in~\va} dt_0 \gamma
 \Rightarrow d E_{in~\va 0} = \gamma d E_{in~\va},
 d E_{in~\va} = \frac{d E_{in~\va 0}}{\gamma}
 \label{dEin0}
\enq
We are now at a position to calculate the variation of the internal energy of a fluid
element:
\beq
 \Delta  d E_{in~\va} = \Delta  \left(\frac{1}{\gamma}\right) d E_{in~\va 0}+  \frac{1}{\gamma} \Delta d E_{in~\va 0} .
 \label{vdEin0}
\enq
Taking into account \ern{thermo2} and \ern{dV0} we obtain:
\beq
 \Delta  d E_{in~\va} = \Delta  \left(\frac{1}{\gamma}\right) d E_{in~\va 0} -  \frac{1}{\gamma} P_0 \Delta dV_{\vec \alpha 0}
 = \Delta  \left(\frac{1}{\gamma}\right) d E_{in~\va 0} -  \frac{1}{\gamma} P_0 \Delta (\gamma dV_{\vec \alpha}).
 \label{vdEin02}
\enq
Thus using the definition of enthalpy given in \ern{thermo2d} we may write:
\beq
 \Delta  d E_{in~\va} =  \Delta  \left(\frac{1}{\gamma}\right) (d E_{in~\va 0} + P_0  dV_{\vec \alpha 0}) -  \ P_0 \Delta dV_{\vec \alpha}=
\Delta  \left(\frac{1}{\gamma}\right) dW_{\va 0} -  \ P_0 \Delta dV_{\vec \alpha}.
 \label{vdEin03}
\enq
We shall now calculate the variation of the volume element. Suppose at a time $t$ the volume of the fluid element labelled by $\va$ is described as:
\beq
 dV_{\va,t} = d^3 x(\va, t)
\label{volelem}
\enq
Using the Jacobian determinant we may relate this to the same element at $t=0$:
\beq
  d^3 x(\va, t)  = J  d^3 x(\va, 0), \qquad
  J \equiv \vec \nabla_0 x_1 \cdot (\vec \nabla_0 x_2 \times \vec \nabla_0 x_3)
\label{volelem2}
\enq
In which $\vec \nabla_0$ is taken with respect to the coordinates of the fluid elements at $t=0$: $\vec \nabla_0 \equiv (\frac{\partial }{\partial x(\va,0)_1},\frac{\partial }{\partial x(\va,0)_2},\frac{\partial }{\partial x(\va,0)_3})$. As both the actual and varied "fluid element" trajectories start at the same point it follows that:
\ber
\Delta dV_{\va,t} &=& \Delta d^3 x(\va, t)  = \Delta J ~ d^3 x(\va, 0)  = \frac{\Delta J}{J} d^3 x(\va, t) = \frac{\Delta J}{J}  dV_{\va,t},
\nonumber \\
(\Delta d^3 x(\va, 0) &=& 0).
\label{volelem3}
\enr
The variation of $J$ can be easily calculated as:
\beq
    \Delta J = \vec \nabla_0 \Delta x_1 \cdot (\vec \nabla_0 x_2 \times \vec \nabla_0 x_3)
    + \vec \nabla_0  x_1 \cdot (\vec \nabla_0 \Delta x_2 \times \vec \nabla_0 x_3)
    + \vec \nabla_0  x_1 \cdot (\vec \nabla_0  x_2 \times \vec \nabla_0 \Delta x_3),
\label{volelem4}
\enq
Now:
\ber
& & \vec \nabla_0 \Delta x_1 \cdot (\vec \nabla_0 x_2 \times \vec \nabla_0 x_3)
= \vec \nabla_0 \xi_1 \cdot (\vec \nabla_0 x_2 \times \vec \nabla_0 x_3)
\nonumber \\
&=& \partial_k \xi_1 \vec \nabla_0 x_k \cdot (\vec \nabla_0 x_2 \times \vec \nabla_0 x_3)
= \partial_1 \xi_1 \vec \nabla_0 x_1 \cdot (\vec \nabla_0 x_2 \times \vec \nabla_0 x_3)=
\partial_1 \xi_1 J.
\nonumber \\
& & \vec \nabla_0  x_1 \cdot (\vec \nabla_0 \Delta x_2 \times \vec \nabla_0 x_3)
= \vec \nabla_0 x_1 \cdot (\vec \nabla_0 \xi_2 \times \vec \nabla_0 x_3)
\nonumber \\
&=& \partial_k \xi_2 \vec \nabla_0 x_1 \cdot (\vec \nabla_0 x_k \times \vec \nabla_0 x_3)
= \partial_2 \xi_2 \vec \nabla_0 x_1 \cdot (\vec \nabla_0 x_2 \times \vec \nabla_0 x_3)=
\partial_2 \xi_2 J.
\nonumber \\
& & \vec \nabla_0  x_1 \cdot (\vec \nabla_0  x_2 \times \vec \nabla_0 \Delta x_3)
= \vec \nabla_0 x_1 \cdot (\vec \nabla_0 x_2 \times \vec \nabla_0 \xi_3)
\nonumber \\
&=& \partial_k \xi_3 \vec \nabla_0 x_1 \cdot (\vec \nabla_0 x_2 \times \vec \nabla_0 x_k)
= \partial_3 \xi_3 \vec \nabla_0 x_1 \cdot (\vec \nabla_0 x_2 \times \vec \nabla_0 x_3)=
\partial_3 \xi_3 J.
\label{volelem5}
\enr
Combining the above results, it follows that:
\beq
    \Delta J = \partial_1 \xi_1 J + \partial_2 \xi_2 J + \partial_3 \xi_3 J =
     \vec \nabla \cdot \vec \xi~ J.
\label{volelem6}
\enq
Which allows us to calculate the variation of the volume of the "fluid element":
\beq
\Delta dV_{\va,t}  = \vec \nabla \cdot \vec \xi ~ dV_{\va,t}.
\label{volelem7}
\enq
And thus the variation of the internal energy given in \ern{vdEin03} is:
\beq
\Delta d E_{in~\vec \alpha} =\Delta  \left(\frac{1}{\gamma}\right) dW_{\va 0}  - P_0 \vec \nabla \cdot \vec \xi ~ dV_{\va,t}.
\label{therm4}
\enq
Taking into account \ern{vgam} this takes the form:
\beq
\Delta d E_{in~\vec \alpha} = - P_{\va 0} \vec \nabla \cdot \vec \xi_\va ~ dV_{\va,t}
- \frac{\gamma_\va  \vec v_\va(t)}{c^2} dW_{\va 0} \cdot \frac{d \vec \xi_\va (t)}{dt}.
\label{therm5}
\enq

The variation  of internal energy is the only novel element with respect to the system of particles scenario described in the previous section, thus the rest of the variation analysis is straight forward. Varying \ern{relaction} we obtain:
\ber
 \Delta d {\cal A}_{\va} &=& \int_{t1}^{t2} \Delta d L_{\va} dt, \qquad \Delta d L_{\va} = \Delta d L_{k\va} + \Delta d L_{i\va} - \Delta d E_{in~\va}
 \nonumber \\
\Delta d L_{k\va} &=&  - d M_{\va} c^2 \Delta \left( \frac{1}{\gamma_\va}\right)
   =  d M_{\va} \gamma_\va  \vec v_\va(t) \cdot \frac{d \vec \xi_\va (t)}{dt},
 \nonumber \\
  \Delta d L_{i\va} &=& d Q_{\va} \left(\Delta \vec A (\vec x_\va (t),t) \cdot \vec v_\va(t)
  + \vec A (\vec x_\va (t),t) \cdot \Delta \vec v_\va(t) \right.
  \nonumber \\
  &-& \left. \Delta \phi(\vec x_\va(t),t) \right).
 \label{varrelparticleactcf}
 \enr

We can now combine the internal and kinetic parts of the varied Lagrangian taking into
account the specific enthalpy definition given in \ern{thermo2e}:
\beq
\Delta d L_{k\va} - \Delta d E_{in~\va}=
 dM_{\va} \gamma_\va (\left(1 + \frac{w_0}{c^2} \right) \vec v_\va(t) \cdot
 \frac{d \vec \xi_\va (t)}{dt} + P_{\va 0} \vec \nabla \cdot \vec \xi_{\va} ~ dV_{\va,t}.
 \label{delLffl}
 \enq
The electromagnetic interaction variation terms are not different than in the low speed (non-relativistic) case, see for example equations A47 and A48 of \cite{Fisherspin3}, and their derivation will not be repeated here:
\beq
 d \vec F_{L\va} \equiv d Q_{\va} \left[ \vec v_{\va} \times \vec B (\vec x_\va (t),t) + \vec E (\vec x_\va  (t),t) \right]
 \label{Lorfl}
 \enq
and:
 \beq
 \Delta d L_{i\va} = \frac{d (d Q_{\va} \vec A (\vec x_\va  (t),t) \cdot \vec \xi_{\va})}{dt} +
  d \vec F_{L\va} \cdot \vec \xi_{\va}.
 \label{delLi4fl}
 \enq
 Introducing the shorthand notation:
 \beq
 \bar \lambda \equiv 1 + \frac{w_0}{c^2}, \qquad \lambda  \equiv \gamma \bar \lambda
 =  \gamma  \left(1 + \frac{w_0}{c^2}\right).
 \label{lambda}
 \enq
  The variation of the action of a relativistic single fluid element is thus:
 \ber
  \Delta d {\cal A}_{\va} &=& \int_{t1}^{t2} \Delta dL_{\va} dt =
  \left. (d M_{\va} \lambda_\va  \vec v_\va (t) + d Q_{\va} \vec A (\vec x_\va  (t),t)) \cdot \vec \xi_{\va} \right|_{t1}^{t2}
  \nonumber \\
   &-& \int_{t1}^{t2}(d M_{\va} \frac{d (\lambda_\va \vec v_\va (t))}{dt}\cdot \vec \xi_{\va} - d \vec F_{L\va} \cdot \vec \xi_{\va} - P_{\va 0} \vec \nabla \cdot \vec \xi_{\va} ~ dV_{\va,t} )  dt.
 \label{delA1fl}
 \enr
The variation of the total action of the fluid is thus:
\ber
\Delta {\cal A} &=& \int_{\va} d {\cal A}_{\vec \alpha} =
\left. \int_{\va} (d M_{\va} \lambda_\va \vec v_\va (t) + d Q_{\va} \vec A (\vec x (\va,t),t)) \cdot \vec \xi_{\va} \right|_{t1}^{t2}
  \nonumber \\
   &-& \int_{t1}^{t2}\int_{\va} (d M_{\va} \frac{d (\lambda_\va \vec v_\va (t))}{dt}\cdot \vec \xi_{\va} - d \vec F_{L\va} \cdot \vec \xi_{\va} - P_{\va 0} \vec \nabla \cdot \vec \xi_{\va} ~ dV_{\va} )  dt.
 \label{varAfl2}
 \enr
 Now according to \ern{dens} we may write:
 \beq
 d M_{\va} = \rho_{\va}~dV_{\va}, \qquad d Q_{\va} = \rho_{c\va}~dV_{\va}
 \label{dvq}
 \enq
using the above relations we may turn the $\va$ integral into a volume integral and thus write
the variation of the fluid action in which we suppress the $\va$ labels:
\beq
\Delta {\cal A} =
\left. \int (\rho \lambda \vec v  + \rho_c \vec A)  \cdot \vec \xi dV \right|_{t1}^{t2}
  - \int_{t1}^{t2}\int (\rho \frac{d (\lambda \vec v)}{dt}\cdot \vec \xi - \vec f_{L} \cdot \vec \xi - P_0 \vec \nabla \cdot \vec \xi)  dV  dt.
 \label{varAfl3}
 \enq
in the above we introduced the Lorentz force density:
\beq
\vec f_{L\va} \equiv \frac{d \vec F_{L\va}}{dV_{\va}} = \rho_{c\va} \left[ \vec v_{\va} \times \vec B (\vec x_\va (t),t) + \vec E (\vec x_\va (t),t) \right].
 \label{Lorflden}
 \enq
Now, since:
\beq
P_0 \vec \nabla \cdot \vec \xi = \vec \nabla \cdot (P_0 \vec \xi) - \vec \xi \cdot \vec \nabla P_0,
\label{Pxi}
 \enq
and using Gauss theorem the variation of the action can be written as:
\ber
\Delta {\cal A} &=&
\left. \int (\rho \lambda \vec v  + \rho_c \vec A)  \cdot \vec \xi dV \right|_{t1}^{t2}
 \nonumber \\
  &-& \int_{t1}^{t2}\left[\int (\rho \frac{d (\lambda \vec v)}{dt}- \vec f_{L} + \vec \nabla P_0)\cdot \vec \xi   dV  - \oint P_0 \vec \xi \cdot d \vec \Sigma \right] dt.
 \label{varAfl4}
 \enr
 It follows that the variation of the action will vanish for a $\vec \xi$ such that $\vec \xi (t1) = \vec \xi (t2) = 0$ and vanishing on a surface encapsulating the fluid, but other than that arbitrary only if the Euler equation for a relativistic charged fluid is satisfied, that is:
 \beq
 \frac{d (\lambda \vec v)}{dt}= -\frac{\vec \nabla P_0}{\rho} +\frac{\vec f_{L}}{\rho}
 \label{Eul1}
 \enq
 for the particular case that the fluid element is made of identical microscopic particles each
 with a mass $m$ and a charge $e$, it follows that the mass and charge densities are proportional to the number density $n$:
 \beq
 \rho = m~n, \quad \rho_c = e~n \Rightarrow \frac{\vec f_{L}}{\rho} = k \left[ \vec v \times \vec B + \vec E  \right], k \equiv \frac{e}{m}
 \label{Eul2}
 \enq
 thus except from the terms related to the internal energy the equation is similar to that of a point particle. For a neutral fluid one obtains the form:
 \beq
 \frac{d (\lambda \vec v)}{dt}= -\frac{\vec \nabla P_0}{\rho}.
 \label{Eul3b}
 \enq
 Some authors prefer to write the above equation in terms of the energy per element
  of the fluid per unit volume in the rest frame which is the sum of the internal energy
  contribution and the rest mass contribution:
  \beq
  e_0  \equiv \rho_0 c^2 + \rho_0 \varepsilon_0.
  \label{eo}
 \enq
 It is easy to show that:
  \beq
  \bar \lambda = 1 + \frac{w_0}{c^2} = \frac{e_0 + P_0}{\rho_0 c^2}.
  \label{eo1}
 \enq
 And using the above equality and some manipulations we may write \ern{Eul3b}
 in a form which is preferable by some authors:
 \beq
(e_0 + P_0) \frac{\gamma }{c^2}\frac{d (\gamma \vec v)}{dt}=
- \vec \nabla P_0 - \frac{\gamma^2}{c^2}\frac{d P_0}{dt} \vec v.
 \label{Eul4b}
 \enq
  In experimental fluid dynamics it is more convenient to describe a fluid in terms of quantities at
 a specific location, rather than quantities associated with unseen infinitesimal "fluid elements". This road leads to the Eulerian description of fluid dynamics and thinking in terms of flow fields rather than in terms of a velocity of "fluid elements" as will be discussed in the next section.

\section{An Eulerian Charged Fluid - the Clebsch Approach}

In this section we follows closely the analysis of \cite{Spflu,Fisherspin3,Fisherspin4} with the modification of taking into account the relativistic corrections, this implies taking into
account an action which is invariant under Lorentz transformations. Let us consider the action:
\ber
 {\cal A} & \equiv & \int {\cal L} d^3 x dt, \qquad
{\cal L}  \equiv  {\cal L}_0 + {\cal L}_2 + {\cal L}_i
\nonumber \\
{\cal L}_0 & \equiv & -\rho (\frac{c^2}{\gamma} + \varepsilon) =
-\rho_0 (c^2 + \varepsilon_0) = - e_0,
\qquad
{\cal L}_2 \equiv   \nu \partial^\nu (\rho_0 u_\nu) - \rho_0 \alpha u_\nu \partial^\nu \beta, \nonumber \\
 {\cal L}_{i} &\equiv & -\rho_{c}  A^\nu  v_\nu, \qquad  v_\nu \equiv \frac{d x_\nu}{dt}.
\label{Lagactionsimpb0}
\enr
In the non relativistic limit we may write:
\beq
{\cal L}_0  \simeq  \rho (\frac{1}{2} v^2 - \varepsilon - c^2)
\label{Lononrel}
\enq
Taking into account that:
\beq
u_\mu = \gamma (c, \vec v)
\label{ustpar}
\enq
and also that $\rho = \gamma \rho_0$ according to \ern{drho0}, it is easy to write the above Lagrangian densities in a space-time formalism:
\beq
 {\cal L}_2 =  \nu [\frac{\partial{\rho}}{\partial t} + \vec \nabla \cdot (\rho \vec v )]
- \rho \alpha \frac{d \beta}{dt}
\qquad
 {\cal L}_{i} = \rho_{c} \left(\vec A \cdot \vec v - \phi \right)
\label{Lagactionsimpb}
\enq
In the Eulerian approach we consider the variational variables to be fields, that functions of
space and time. We have two such variational variables the vector velocity field $\vec v (\vec x,t)$ and density scalar field $\rho (\vec x,t)$. The conservation of quantities such as
the label of the fluid element, mass, charge and entropy are dealt by introducing Lagrange multipliers $\nu,\alpha$ in such a way that the variational principle will yield the following \ens:
\ber
& & \frac{\partial{\rho}}{\partial t} + \vec \nabla \cdot (\rho \vec v ) = 0
\nonumber \\
& & \frac{d \beta}{dt} = 0
\label{lagmulb}
\enr
Provided $\rho$ is not null those are just the continuity equation which ensures mass conservation and the conditions that $\beta$ is comoving and is thus a label.
Let us now calculate the variation with respect to $\beta$, this will lead us to the following results:
\ber
\delta_{\beta} A & = & \int d^3 x dt \delta \beta
[\frac{\partial{(\rho \alpha)}}{\partial t} +  \vec \nabla \cdot (\rho \alpha \vec v)]
\nonumber \\
 & - & \oint d \vec S \cdot \vec v \rho \alpha \delta \beta -\int d \vec \Sigma \cdot \vec v \rho \alpha [\delta \beta]
 - \int d^3 x \rho \alpha \delta \beta |^{t_1}_{t_0}
\label{delActionchib}
\enr
Hence choosing $\delta \beta$ in such a way that the temporal and
spatial boundary terms vanish (this includes choosing $\delta \beta$ to be continuous on the cut if one needs
to introduce such a cut) in the above integral will lead to
the equation:
\beq
\frac{\partial{(\rho \alpha)}}{\partial t} +  \vec \nabla \cdot (\rho \alpha \vec v) =0
\enq
Using the continuity \ern{lagmulb} this will lead to the equation:
\beq
\frac{d \alpha}{dt} = 0
\label{alphacon}
\enq
Hence for $\rho \neq 0$ both $\alpha$ and $\beta$ are comoving coordinates.
This is why in the Eulerian approach we are obliged
to add the Lagrangian density ${\cal L}_2$. The specific internal energy $\varepsilon_0$ defined
in \ern{specifint} is dependent on the thermodynamic properties of the specific fluid. That is it generally depends through a given "equation of state" on the density and specific entropy. In our case we shall assume a barotropic fluid, that is a fluid in which $\varepsilon_0(\rho_0)$ is a function of the density $\rho_0$ only. Other functions connected to the electromagnetic interaction such as the potentials $\vec A, \phi$ are assumed given function of coordinates and are not varied.
Another simplification which we introduce is the assumption the fluid element is made of microscopic
particles having a given mass $m$ and a charge $e$, in this case it follows from \ern{Eul2} that:
\beq
 \rho_c = k \rho.
 \label{rhoc}
 \enq
Let us now take the variational derivative with respect to the density $\rho$, we obtain:
\ber
\delta_{\rho} A & = & \int d^3 x dt \delta \rho
[-\frac{c^2}{\gamma} - w_0 \frac{\delta \rho_0}{\delta \rho}  - \frac{\partial{\nu}}{\partial t} -  \vec v \cdot \vec \nabla \nu+
k(\vec A \cdot \vec v - \phi)]
\nonumber \\
 & + & \oint d \vec S \cdot \vec v \delta \rho  \nu +\int d \vec \Sigma \cdot \vec v \delta \rho  [\nu] +
  \int d^3 x \nu \delta \rho |^{t_1}_{t_0}
\label{delActionrhob}
\enr
Or as:
\ber
\delta_{\rho} A & = & \int d^3 x dt \delta \rho
[-\frac{c^2+w_0}{\gamma}   - \frac{\partial{\nu}}{\partial t} -  \vec v \cdot \vec \nabla \nu+
k(\vec A \cdot \vec v - \phi)]
\nonumber \\
 & + & \oint d \vec S \cdot \vec v \delta \rho  \nu +\int d \vec \Sigma \cdot \vec v \delta \rho  [\nu] +
  \int d^3 x \nu \delta \rho |^{t_1}_{t_0}
\label{delActionrhob2}
\enr
in which $w_0 = \frac{\partial(\rho_0 \varepsilon_0 )}{\partial \rho_0}$ is the specific enthalpy in the rest frame of the fluid element (see \ern{thermo2f}). Hence provided that $\delta \rho$ vanishes on the boundary of the domain, on the cut and in initial and final times the following \eqn must be satisfied:
\beq
\frac{d \nu}{d t} = \frac{\partial \nu}{\partial t} + \vec v \cdot \vec \nabla \nu = -\frac{c^2+w_0}{\gamma} + k(\vec A \cdot \vec v - \phi)
\label{nueqb}
\enq
In the above we notice that taking a time derivative for a fixed label $\va$ (also known as a material derivative) of any quantity $g$ takes the form:
\beq
\frac{d g (\va,t) }{d t}  = \frac{d g (\vec x (\va,t),t) }{d t}
= \frac{\partial g}{\partial t} + \frac{d \vec x }{d t} \cdot \vec \nabla g
= \frac{\partial g}{\partial t} + \vec v \cdot \vec \nabla g
\label{materderiv}
\enq
once $g$ is considered to be a field dependent on $\vec x,t$.

Finally Let us take an arbitrary variational derivative of the above action with
respect to $\vec v$, taking into account that:
\beq
\delta_{\vec v} \frac{1}{\gamma} =  - \gamma \frac{\vec v \cdot \delta \vec v}{c^2}
\label{delvgam}
\enq
This will result in:
\ber
\delta_{\vec v} A & = & \int d^3 x dt \rho \delta \vec v \cdot
[\gamma \vec v  - \frac{w_0}{\rho} \frac{\delta \rho_0}{\delta \vec v}- \vec \nabla \nu - \alpha \vec \nabla \beta + k \vec A]
\nonumber \\
 & + & \oint d \vec S \cdot \delta \vec v \rho \nu + \int d \vec \Sigma \cdot \delta \vec v \rho [\nu].
\label{delActionvb}
\enr
However:
\beq
\frac{\delta \rho_0}{\delta \vec v}= \rho \frac{\delta \frac{1}{\gamma}}{\delta \vec v}
=  - \rho \gamma \frac{\vec v }{c^2}
\label{delrho0v}
\enq
Taking in account the definition of $\lambda$ (see \ern{lambda}), we thus have:
\ber
\delta_{\vec v} A & = & \int d^3 x dt \rho \delta \vec v \cdot
[\lambda \vec v  - \vec \nabla \nu - \alpha \vec \nabla \beta + k \vec A]
\nonumber \\
 & + & \oint d \vec S \cdot \delta \vec v \rho \nu + \int d \vec \Sigma \cdot \delta \vec v \rho [\nu].
\label{delvActionvrel}
\enr
the above boundary terms contain integration over the external boundary $\oint d \vec S$ and an
integral over the cut $\int d \vec \Sigma$ that must be introduced in case that $\nu$ is not single
valued, more on this case in later sections.
The external boundary term vanishes; in the case of astrophysical flows for which $\rho=0$ on the free flow
boundary, or the case in which the fluid is contained in a vessel which induces a no flux boundary condition $\delta \vec v \cdot \hat n =0$
($\hat n$ is a unit vector normal to the boundary). The cut "boundary" term vanish when the velocity field varies only parallel
to the cut that is it satisfies a Kutta type condition. If the boundary terms vanish  $\vec v$ must have the following form:
\beq
\lambda \vec v = \alpha \vec \nabla \beta + \vec \nabla \nu -  k \vec A
\label{vformb}
\enq
this is a generalization of Clebsch representation of the flow field (see for example \cite{Eckart}, \cite[page 248]{Lamb H.}) for a relativistic charged flow.

\subsection{Euler's equations}
\label{Eulerequations}

We shall now show that a velocity field given by \ern{vformb}, such that the
functions $\alpha, \beta, \nu$ satisfy the corresponding equations
(\ref{lagmulb},\ref{nueqb},\ref{alphacon}) must satisfy Euler's equations.
Let us calculate the material derivative of $\lambda \vec v$:
\beq
\frac{d(\lambda \vec v)}{dt} = \frac{d\vec \nabla \nu}{dt}  + \frac{d\alpha}{dt} \vec \nabla \beta +
 \alpha \frac{d\vec \nabla \beta}{dt} - k \frac{d\vec A}{dt}
\label{dvform12b}
\enq
It can be easily shown that:
\ber
\frac{d\vec \nabla \nu}{dt} & = & \vec \nabla \frac{d \nu}{dt}- \vec \nabla v_n \frac{\partial \nu}{\partial x_n}
 = \vec \nabla \left(-\frac{c^2 + w_0}{\gamma}  + k \vec A \cdot \vec v -k \phi\right)- \vec \nabla v_n \frac{\partial \nu}{\partial x_n}
 \nonumber \\
 \frac{d\vec \nabla \beta}{dt} & = & \vec \nabla \frac{d \beta}{dt}- \vec \nabla v_n \frac{\partial \beta}{\partial x_n}
 = - \vec \nabla v_n \frac{\partial \beta}{\partial x_n}
  \label{dnablab}
\enr
In which $x_n$ is a Cartesian coordinate and a summation convention is assumed. Inserting the result from equations
(\ref{dnablab}) into \ern{dvform12b} yields:
\ber
& & \frac{d (\lambda \vec v)}{dt} = - \vec \nabla v_n (\frac{\partial \nu}{\partial x_n} +
 \alpha \frac{\partial \beta}{\partial x_n} ) + \vec \nabla \left(-\frac{c^2 + w_0}{\gamma} + k \vec A \cdot \vec v -k \phi \right) - k \frac{d\vec A}{dt}
 \nonumber \\
&=& - \vec \nabla v_n (\lambda v_n + k A_n) + \vec \nabla (-\frac{c^2 + w_0}{\gamma} + k \vec A \cdot \vec v -k \phi) - k \partial_t \vec A - k (\vec v \cdot \vec \nabla) \vec A
\nonumber \\
&=& - \frac{1}{\gamma} \vec \nabla w_0 + k \vec E + k (v_n \vec \nabla A_n - v_n \partial_n \vec A),
\label{dvform2bb}
\enr
in the above we have used the electric field defined in \ern{electric}. We notice that
according to \ern{magnetic}:
 \beq
 (v_n \vec \nabla A_n - v_n \partial_n \vec A)_l  = v_n (\partial_l  A_n -\partial_n  A_l)=
\epsilon_{lnj} v_n B_j = (\vec v \times \vec B)_l,
\label{EB3}
 \enq
 Hence we obtain the Euler equation of a charged relativistic fluid in the form:
 \beq
 \frac{d (\lambda \vec v)}{dt}= - \frac{1}{\gamma}\vec \nabla w_0 + k \left[ \vec v \times \vec B + \vec E  \right]
 = - \frac{1}{\rho} \vec \nabla P_0 + k \left[ \vec v \times \vec B + \vec E  \right],
  \label{Eul5}
 \enq
 since  (see \ern{thermo2g}):
 \beq
 \vec \nabla w_0 = \frac{\partial w_0 }{\partial \rho_0} \vec \nabla \rho_0 =
  \frac{1}{\rho_0} \frac{\partial P_0}{\partial \rho_0} \vec \nabla \rho_0 = \frac{1}{\rho_0} \vec \nabla P_0.
  \label{Eul6}
 \enq
  The above equation is identical to \ern{Eul1} and thus proves that the Euler equations can be derived from the action given in \ern{Lagactionsimpb} and hence
all the equations of charged fluid dynamics can be derived from the above action
without restricting the variations in any way.

\subsection{Simplified action}
\label{simpact}
The reader of this paper might argue that the authors have introduced unnecessary complications
to the theory of relativistic fluid dynamics by adding three  more functions $\alpha,\beta,\nu$ to the standard set
$\vec v,\rho$. In the following we will show that this is not so and the action given in \ern{Lagactionsimpb0} in a form suitable for a pedagogic presentation can indeed be simplified. It is easy to show that defining a four dimensional Clebsch four vector:
\beq
v_{C}^{\mu} \equiv \alpha \partial^\mu \beta + \partial^\mu \nu= (\frac{1}{c}(\alpha \partial_t \beta + \partial_t \nu), \alpha \vec \nabla \beta + \vec \nabla \nu)
= (\frac{1}{c}(\alpha \partial_t \beta + \partial_t \nu), \vec v_C)
\label{vCmu}
\enq
and a four dimensional electromagnetic Clebsch four vector:
\beq
v_{E}^{\mu} \equiv v_{C}^{\mu} + k A^\mu
= (\frac{1}{c}(\alpha \partial_t \beta + \partial_t \nu + k \phi), \vec v_C - k \vec A).
\label{vEmu}
\enq
It follows from \ern{nueqb} and \ern{vformb} that:
\beq
v_{\mu} = -\frac{v_{E\mu}}{\lambda} \Rightarrow \vec v = \frac{\vec v_E}{\lambda} .
\label{vCmu2}
\enq
Eliminating $\vec v$ the Lagrangian density appearing in \ern{Lagactionsimpb} can be written (up to surface terms) in the compact form:
\beq
{\cal L} [\rho_0,\alpha,\beta,\nu] = \rho_0 \left[c \sqrt{v_{E \mu} v_{E}^{\mu}}- \varepsilon_0 - c^2 \right]
\label{Lagactionsimpb4}
\enq
This Lagrangian density will yield the four \eqs
(\ref{lagmulb},\ref{alphacon},\ref{nueqb}), after those equations are solved we can insert the potentials $\alpha,\beta,\nu$ into \ern{vformb} to obtain the physical velocity $\vec v$.
Hence, the general charged relativistic barotropic fluid dynamics problem is changed such that instead of solving the Euler and continuity equations we need to solve an alternative set which can be derived from the Lagrangian density $\hat {\cal L}$.

\section {Conclusion}

The current work which is a continuation of previous studies \cite{Spflu,Fisherspin,Fisherspin2}
in which we demonstrate how Pauli's spinor can be interpreted in terms of spin fluid  using a generalized Clebsch form which is modified to include the electromagnetic vector potential affecting a charged fluid. The theory is described by an action and a variational principle and the fluid equations are derived as the extrema of the action. The similarities as well as the pronounced differences with barotropic fluid dynamics were discussed.

 A fundamental obstacle to the fluid interpretation of quantum mechanics still exist. This is related to the origin of thermodynamic quantities which are part of fluid mechanics in the quantum context. For classical fluid the thermodynamic internal energy implies that a fluid element is not a point particle but has internal structure. In standard thermodynamics notions as specific enthalpy,  pressure and temperature are  derived from the specific internal energy equation of state.
 The internal energy is a required component of any Lagrangian density attempting to depict a fluid.
 The unique form of the internal energy can be derived in principle relying on the basis of the atoms and  molecules from which the fluid is composed and their interactions using statistical physics. However, the quantum fluid has no such microscopic structure
 and yet analysis of both the spin less \cite{Madelung,Complex} and spin \cite{Spflu} quantum fluid shows that terms analogue to internal energies appear. Thus one is forced to ask where do those internal energies originate, surely the quantum fluid is devoid of a microscopic sub structure as this will defy the empirically supported conception of the electron as a point particle. The answer to this inquiry originated from measurement theory \cite{Fisher}. Fisher information a basic concept of measurement theory is a measure of the quality of the  measurement. It was shown that this concept is proportional to the internal energy of Schr\"{o}dinger's spin-less electron which is essentially a theory of a potential flow which moves under the influence of electromagnetic fields and fisher information forces. Fisher information  can also explain  most parts of the internal energy of an electron with spin. This puts (Fisher) information as a fundamental force of  nature, which has the same status as electromagnetic forces in the quantum mechanical level of reality. Indeed, according to  Anton Zeilinger's recent remark to the press, it is quantum mechanics that demonstrates that information is more fundamental than space-time.

We have highlighted the similarities between the variational principles of Eulerian fluid mechanics and both Schr\"{o}dinger's and Pauli's quantum mechanics as opposed to classical mechanics. The former have only linear time derivatives of degrees of freedom while that later have quadratic time derivatives. The former contain terms quadratic in the vector potential $\vec A$ while the later contain only linear terms.

While the analogies between spin fluid dynamics classic Clebsch fluid dynamics are quite convincing still there are terms in spin fluid dynamics that lack classical interpretation. It was thus suggested that those term originate from a relativistic Clebsch theory which was the main motivation to the current paper. Indeed following the footsteps of the pervious papers \cite{Fisherspin,Fisherspin2}
we may replace the internal energy in \ern{Lagactionsimpb4} with a Lorentz invariant Fisher information term to obtain a new Lagrangian density of relativistic quantum mechanics of a particle with spin:
\beq
{\cal L} [\rho_0,\alpha,\beta,\nu] = \rho_0 \left[c \sqrt{v_{E \mu} v_{E}^{\mu}} - c^2 \right]
- \frac{\hbar^2}{2 m} \partial^\mu a_0 \partial_\mu a_0, \qquad a_0 \equiv \sqrt{\frac{\rho_0}{m}}.
\label{Lagquantum}
\enq
in the above $m$ is the particle's mass and $\hbar$ is Planck's constant divided by $2 \pi$.

A side benefit of the above work is the ability to canonically derive the stress energy tensor of
a relativistic fluid.

As the current paper is of limited scope, we were not able to compare the above lagrangian with its low speed limit and derive the relevant quantum equation, hopefully this will be done in a following more expanded paper.

Not less important is the comparison between the fluid route to relativistic quantum mechanics and
the more established route of the Dirac equation, this certainly deserve and additional paper which I hope to compose in the near future.


\begin {thebibliography}9

 \bibitem{Kant}
Kant, I. (1781). Critik der reinen Vernunft.
\bibitem{Bohm} D. Bohm, {\it Quantum Theory} (Prentice Hall, New York, 1966)  section 12.6
 \bibitem{Holland}
P.R. Holland {\it The Quantum Theory of Motion} (Cambridge University Press, Cambridge, 1993)
 \bibitem{DuTe}
D. Durr \& S. Teufel {\it Bohmian Mechanics: The Physics and Mathematics of Quantum Theory} (Springer-Verlag, Berlin Heidelberg, 2009)
\bibitem{Madelung}
E. Madelung, Z. Phys., {\bf 40} 322 (1926)
\bibitem {Complex}
R. Englman and A. Yahalom "Complex States of Simple Molecular Systems"
a chapter of the volume "The Role of Degenerate States in Chemistry" edited by M.
Baer and G. Billing in Adv. Chem. Phys. Vol. 124 (John Wiley \& Sons 2002). [Los-Alamos Archives physics/0406149]
\bibitem{Pauli}
W. Pauli (1927) Zur Quantenmechanik des magnetischen Elektrons Zeitschrift f\"{u}r Physik (43) 601-623
\bibitem {EYB1}
R. Englman, A.Yahalom and M. Baer, J. Chem. Phys.{109} 6550 (1998)
\bibitem {EYB2}
R. Englman, A. Yahalom and M. Baer, Phys. Lett. A {\bf 251} 223 (1999)
\bibitem {EY1}
R. Englman and A. Yahalom, Phys. Rev. A {\bf 60} 1802 (1999)
\bibitem {EYB3}
R. Englman, A.Yahalom and M. Baer, Eur. Phys. J. D {\bf 8} 1 (2000)
\bibitem {EY2}
R. Englman and A.Yahalom, Phys. Rev. B {\bf 61} 2716 (2000)
\bibitem {EY3}
R. Englman and A.Yahalom, Found. Phys. Lett. {\bf 13} 329 (2000)
\bibitem {EY4}
R. Englman and A.Yahalom, {\it The Jahn Teller Effect: A Permanent Presence
in the Frontiers of Science} in M.D. Kaplan and G. Zimmerman (editors),
{\it Proceedings of the NATO Advanced Research
Workshop, Boston, Sep. 2000}  (Kluwer, Dordrecht, 2001)
\bibitem {BE2}
M. Baer and R. Englman, Chem. Phys. Lett. {\bf 335} 85 (2001)
\bibitem {MBEY}
A. Mebel, M. Baer, R. Englman and A. Yahalom, J.Chem. Phys. {\bf 115} 3673 (2001)
\bibitem {EYBM4MCI}
R. Englman \& A. Yahalom, "Signed Phases and Fields Associated with Degeneracies" Acta Phys. et Chim., 34-35, 283 (2002). [Los-Alamos Archives - quant-ph/0406194]
\bibitem {EY5}
R. Englman, A. Yahalom and M. Baer,"Hierarchical Construction of Finite Diabatic Sets, By Mathieu Functions", Int. J. Q. Chemistry, 90, 266-272 (2002). [Los-Alamos Archives -physics/0406126]
\bibitem {EY6}
R. Englman, A. Yahalom, M. Baer and A.M. Mebel "Some Experimental. and Calculational Consequences of Phases in Molecules with Multiple Conical Intersections" International Journal of Quantum Chemistry, 92, 135-151 (2003).
\bibitem {EY7}
R. Englman \& A. Yahalom, "Phase Evolution in a Multi-Component System", Physical Review A, 67, 5, 054103-054106 (2003). [Los-Alamos Archives -quant-ph/0406195]
\bibitem {EY8}
R. Englman \& A. Yahalom, "Generalized "Quasi-classical" Ground State of an Interacting Doublet" Physical Review B, 69, 22, 224302 (2004). [Los-Alamos Archives - cond-mat/0406725]
\bibitem {Spflu}
A. Yahalom "The Fluid Dynamics of Spin".  Molecu\-lar Physics, Published online: 13 Apr 2018.\\
 http://dx.doi.org/10.1080/00268976.2018.1457808\\ (arXiv:1802.09331 [physics.flu-dyn]).
\bibitem{Clebsch1}
Clebsch, A., Uber eine allgemeine Transformation der hydrodynamischen Gleichungen.
{\itshape J.~reine angew.~Math.}~1857, {\bf 54}, 293--312.
\bibitem{Clebsch2}
Clebsch, A., Uber die Integration der hydrodynamischen Gleichungen.
{\itshape J.~reine angew.~Math.}~1859, {\bf 56}, 1--10.
\bibitem {Davidov}
B. Davydov\index{Davydov B.}, "Variational principle and canonical
equations for an ideal fluid," {\it Doklady Akad. Nauk},  vol. 69,
165-168, 1949. (in Russian)
\bibitem {Eckart}
C. Eckart\index{Eckart C.}, "Variation\index{variational
principle} Principles of Hydrodynamics \index{hydrodynamics},"
{\it Phys. Fluids}, vol. 3, 421, 1960.
\bibitem {Bertherton}
F.P. Bretherton "A note on Hamilton's principle for perfect fluids," Journal of Fluid Mechanics / Volume 44 / Issue 01 / October 1970, pp 19 31 DOI: 10.1017/S0022112070001660, Published online: 29 March 2006.
\bibitem{Herivel}
J. W. Herivel  Proc. Camb. Phil. Soc., {\bf 51}, 344 (1955)
\bibitem{Serrin}
J. Serrin, {\it \lq Mathematical Principles of Classical Fluid
Mechanics'} in {\it Handbuch der Physik}, {\bf 8}, 148 (1959)
\bibitem{Lin}
C. C. Lin , {\it \lq Liquid Helium'} in {\it Proc. Int. School Phys. XXI}
(Academic Press)  (1963)
\bibitem{Seliger}
R. L. Seliger  \& G. B. Whitham, {\it Proc. Roy. Soc. London},
A{\bf 305}, 1 (1968)
\bibitem{LynanKatz}
D. Lynden-Bell and J. Katz "Isocirculational Flows and their Lagrangian and Energy principles",
Proceedings of the Royal Society of London. Series A, Mathematical and Physical Sciences, Vol. 378,
No. 1773, 179-205 (Oct. 8, 1981).
\bibitem{KatzLyndeb}
J. Katz \& D. Lynden-Bell 1982,{\it Proc. R. Soc. Lond.} {\bf A 381} 263-274.
\bibitem{YahLyndeb}
Asher Yahalom and Donald Lynden-Bell "Variational Principles for Topological Barotropic Fluid Dynamics" ["Simplified Variational Principles for Barotropic Fluid Dynamics" Los-Alamos Archives - physics/ 0603162] Geophysical \& Astrophysical Fluid Dynamics. 11/2014; 108(6). DOI: 10.1080/03091929.2014.952725.
\bibitem{Weinberg}
Weinberg, S. \emph{Gravitation and Cosmology: Principles and Applications of the General Theory of Relativity};
John~Wiley \& Sons, Inc.:  Hoboken, NJ, USA, 1972.
\bibitem{MTW}
Misner, C.W.; Thorne, K.S.; Wheeler, J.A. \emph{Gravitation};  W.H. Freeman \& Company:  New York, NY, USA, 1973.
\bibitem{Padma}
Padmanabhan, T. \emph{Gravitation: Foundations and Frontiers}; Cambridge University Press: Cambridge, UK, 2010.
\bibitem {Goldstein}
H. Goldstein , C. P. Poole Jr. \& J. L. Safko, Classical Mechanics, Pearson; 3 edition (2001).
\bibitem{LyndenB}
Lynden-Bell, D., Consequences of one spring researching with Chandrasekhar. {\itshape Current Science} 1996, {\bf 70}(9), 789--799.
\bibitem{Lamb H.}
Lamb, H., {\itshape Hydrodynamics}, 1945 (New York: Dover Publications).
\bibitem{Moff2}
H.K. Moffatt, \index{Moffatt H.K.} "The degree of knottedness of tangled vortex lines," {\it J. Fluid Mech.}, vol. 35, 117, 1969.
\bibitem{Schrodinger} E. Schr\"{o}dinger, Ann. d. Phys. {\bf 81} 109 (1926).
 English translation appears in E. Schr\"{o}dinger, {\it Collected Papers in Wave
 Mechanics} (Blackie and Sons, London, 1928) p. 102
  \bibitem{Fisher}
 R. A. Fisher {\it Phil. Trans. R. Soc. London} {\bf 222}, 309.
\bibitem{YaFisher}
A. Yahalom "Gravity and the Complexity of Coordinates in Fisher Information" International Journal of Modern Physics D, Vol. 19, No. 14 (2010) 2233-2237, \copyright World Scientific Publishing Company DOI: 10.1142/S0218271810018347.
 \bibitem {Mandel}
L. Mandel and E. Wolf, {\it Optical Coherence and Quantum Optics} (University
Press, Cambridge, 1995) section 3.1
\bibitem {Fisherspin}
A. Yahalom "The Fluid Dynamics of Spin - a Fisher Information Perspective" arXiv:1802.09331v2 [cond-mat.] 6 Jul 2018. Proceedings of the Seventeenth Israeli - Russian Bi-National Workshop 2018 "The optimization of composition, structure and properties of metals, oxides, composites, nano and amorphous materials".
\bibitem {Frieden}
B. R. Frieden {\it Science from Fisher Information: A Unification} (Cambridge University Press, Cambridge, 2004)
\bibitem {Fisherspin2}
Asher Yahalom "The Fluid Dynamics of Spin - a Fisher Information Perspective and Comoving Scalars" Chaotic Modeling and Simulation (CMSIM) 1: 17-30, 2020.
\bibitem {Fisherspin3}
Yahalom, A. Fisher Information Perspective of Pauli's Electron. Entropy 2022, 24, 1721. https://doi.org/10.3390/e24121721.
\bibitem {Fisherspin4}
Yahalom, A. (2023). Fisher Information Perspective of Pauli’s Electron. In: Skiadas, C.H., Dimotikalis, Y. (eds) 15th Chaotic Modeling and Simulation International Conference. CHAOS 2022. Springer Proceedings in Complexity. Springer, Cham.
https://doi.org/10.1007/978-3-031-27082-6\_26
\end {thebibliography}
\end {document}